**Engineering Sustainability**

# The environmental value of transport infrastructure in the UK: an EXIOBASE analysis

**Nikolaos Kalyviotis** Dipl-Ing, MSc, MSPM, MBA, PhD
Assistant Professor of Planning, Management and Evaluation of Programs, Investments and Projects, Department of Planning and Regional Development, School of Engineering, University of Thessaly, Volos, Greece (Orcid:0009-0006-5658-4262) (corresponding author: nkalyviotis@uth.gr)

**Christopher D. F. Rogers** BSc, PhD, CEng, MICE, FCIHT, FISTT, SFHEA, Eur Ing
Professor Geotechnical Engineering, Department of Civil Engineering, University of Birmingham, Birmingham, UK (Orcid:0000-0002-1693-1999)

**Geoffrey J. D. Hewings** BA, MA, PhD
Emeritus Professor of Urban and Regional Planning, Economics and Geography and Regional Science, Regional Economics Applications Laboratory, University of Illinois at Urbana–Champaign, Urbana, IL, USA (Orcid:0000-0003-2560-3273)

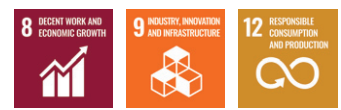

Five life cycle assessment (LCA) methods to calculate a project's environmental value are described: (*a*) process-based, (*b*) hybrid, (*c*) pseudo, (*d*) simplified, and (*e*) parametric. This paper discusses in detail and compares the two methods with the least inherent uncertainty: process-based LCA (a bottom-up methodology involving mapping and characterising all processes associated with all life cycle phases of a project) and a hybrid LCA (the EXIOBASE analysis, which incorporates top-down economic input–output analysis and is a wider sector-by-sector approach). The 'bottom-up' nature of process-based LCA, which quantifies the environmental impacts for each process in all life cycle phases of a project, is particularly challenging when applied to the evaluation of infrastructure as a whole. Conversely, combining the environmental impact information provided in EXIOBASE tables with the corresponding input–output tables allows decision makers to more straightforwardly choose to invest in infrastructure that supports positive environmental outcomes. Employing LCA and a bespoke model using Pearson's correlation coefficient to capture environmental interdependencies between the transport sector and the other four 'economic infrastructures' showed the transport and energy sectors to be most closely linked. Both integrated planning and innovative technologies are needed to radically reduce adverse environmental impacts and enhance sustainability across transport, waste, water, and communication sectors.

## 1. Introduction

Modern society is wholly reliant on civil infrastructure, both to facilitate civilised life and to do so while supporting a move to a more sustainable, resilient, and liveable world—it is a critical determinant of how well society is progressing. The immediate political perspective is that economic activities facilitated by energy, water, waste, transportation, and communication infrastructures (which are indeed termed the 'economic infrastructures') are essential for societal prosperity. However, it is also important to recognise that the natural environment plays a crucial role in supporting these infrastructures and that it is vital for both economic prosperity and human life to flourish. Consequently, protection and enhancement of the natural environment should be prioritised by governments and decision makers when advancing infrastructure engineering, which has the potential to deliver multiple forms of economic, social, and environmental value if systemically designed with foresight in mind.

To achieve this, it is important for decision makers to recognise the diverse perspectives on environmental value: decisions should not be based on a single definition or metric for the measurement of benefits. A further critical consideration for decision makers is the context in which the assessment of the environmental value occurs. Properly defining the appropriate context will help decision makers to understand the benefits and limitations of alternative environmental value analysis methods. Here, it is helpful to distinguish between environmental impacts (the outcomes deriving from design choices when implemented) and value (the components that feature in 'business models', which balance all value realised or compromised by a system intervention—a changed policy, operational practice, a new construction project or whatever; see Bouch *et al*., 2018). Business models should seek to capture all forms of value arising from a project: direct and indirect economic, social, environmental, and cultural value (Rogers *et al*., 2023).

This research addresses these issues and aims to broaden the context in which environmental decisions are made. The scope of this paper includes a discussion of five life cycle assessment (LCA) methods, with a detailed focus on the two methods with the least uncertainty: process-based LCA and hybrid LCA (EXIOBASE). The research reported herein applies these methods to the transport infrastructure sector in the UK to provide a comprehensive sector-wide analysis while investigating the interdependencies between the transport system and four other economic infrastructure systems (energy, waste, water, and communication). The aim is to enhance sustainability through integrated planning.

First, it defines environmental value in relation to civil infrastructure. It then provides a general discussion on environmental value and delves into the specific challenges associated with assessing the environmental impact of civil infrastructures, moving from a broad perspective to a detailed literature review. The research aims to establish the most suitable method for assessing the environmental impacts (hence value gained or lost) by comparing two traditional assessment methods: process-based LCAs (e.g. see Jones *et al*., 2017) and environmentally extended multi-regional input–output (EE MRIO) tables, such as EXIOBASE 3 (Stadler *et al*., 2018).

The starting hypothesis is that a hybrid LCA is a more effective approach for infrastructure and policy decision making, particularly when decisions are made at a higher level and a top-down analysis is employed (e.g. focusing on the civil infrastructure of society as a whole rather than on specific infrastructure projects; see Kalyviotis, 2022; Kalyviotis *et al*., 2018). However, it is appreciated that each method has its own value for different applications, and the benefits and limitations of each will be discussed in detail herein.

Infrastructure is defined as a 'large-scale physical resource made by humans for public consumption' (Frischmann, 2012: p. 3) essential for meeting the needs of people and ensuring the functionality of the economy. However, infrastructure projects usually affect ecosystems by altering, or even destroying, species' habitats (Lederman and Wachs, 2014). These impacts are typically addressed on a project-by-project basis using tools such as the habitat conservation plan (Lederman and Wachs, 2014). Habitat conservation plans are developed during the planning phase of an infrastructure project, focusing on the effect of the construction project (Lederman and Wachs, 2014) rather than on the systemic impacts of the infrastructure. At a systemic scale, new infrastructure will interact with existing infrastructures and the natural environment, thereby amplifying the scope of its environmental impacts.

The novelty of this research lies in its comprehensive comparative analysis of process-based LCA and a hybrid LCA method utilising EXIOBASE. The study focuses on evaluating the uncertainties and the applicability of these methods to large-scale infrastructure projects. Uniquely, it applies the EXIOBASE method to assess the environmental impact of the entire transport infrastructure sector in the UK rather than individual projects. In addition, a novel model employing Pearson's correlation coefficient has been developed to capture environmental interdependencies between the transport system and other economic infrastructure systems. This approach is therefore innovative within the context of environmental impact assessment.

## 2. Literature review

### 2.1 Perspectives on environmental value

An important dimension of the literature on environmental value relates to the utility of nature. Environmental value has a multilateral nature (Stamatopoulos, 2021). In the late 1990s, the notion emerged that environmental value is often not appreciated until it is disrupted or lost (Daily, 1997). For instance, the importance of forests in the hydrological cycle of an ecosystem became apparent only after deforestation led to flash flooding, significant erosion, and other negative effects (Daily, 1997). These adverse effects impact human, animal, and plant life within an ecosystem. This literature highlights both the utility value and the environmental value lost when ecosystems are degraded due to human activity. More recent discussions concluded that environmental value encompasses the human ability to use nature kindly and responsibly, contributing equitably to quality of life for present and future generations through a cross-disciplinary collaboration between societies and human activities (Reser and Bentrupperbäumer, 2005).

Currently, the debate on environmental value centres on whether it should focus on the natural systems (e.g. nature, ecosystems, animals, etc.) or on the benefits derived from these systems that support growth and human life (Chan *et al*., 2016). These ideas relate to the tension between what is termed 'strong and weak sustainability practices' in business and government. Strong sustainability posits that humans are part of nature, fostering a partnership-style relationship, whereas weak sustainability suggests that humans control nature, implying a relationship of human dominance over the environment (Landrum, 2018).

Although both strong and weak sustainability theories aim to reduce negative impacts on the environment, decision makers exert a strong influence on the sustainability measures adopted. How environmental values are viewed, understood, and incorporated into decision making is influenced by societal perspectives on natural systems. Chan *et al*. (2016) argue that the environment has three dimensions of value: intrinsic, instrumental, and relational. Intrinsic value refers to the value that the natural systems have for the rest of the ecosystem, excluding ('independent of people'), while instrumental value is the value that 'brings pleasure or satisfaction' to people and relational value exists in the personal and social relationships that humans have with nature (Chan *et al*., 2016).

Both strong and weak sustainability theories engage with the instrumental value of natural systems (Landrum, 2018) since they extend to the delivery of ecosystem services (or satisfying needs). The weak sustainability model seeks to extract materials and resources from nature to a degree that minimises harm to the environment (Landrum, 2018), while the strong sustainability model focuses on extracting, recycling, and replenishing materials, resources, and energy in nature (Landrum, 2018). One challenge with sustainability theories is that they typically do not directly engage with the intrinsic and relational values of nature. The environment has an inherent value that should be protected, independent of human interaction or need (Gómez-Baggethun and Ruiz-Pérez, 2011). Culturally, spiritually, socially, and mentally, humans derive significant value from the environment in ways that it is very challenging to quantify (Chan *et al*., 2016).

Herein lies an important difference between economic and environmental value; economic value can be transferred and replaced, but the intrinsic, instrumental, and relational value of the environment cannot (Chan *et al*., 2016). When studying interdependencies, economic value in one sector (e.g. infrastructure) can be transferred to another sector or industry, but environmental loss of value in one sector or industry cannot be rectified by environmental improvements in another sector or industry. Damage to an ecosystem by infrastructure systems is often permanent and has ripple effects on the surrounding environment and, consequently, on society.

The Gaia hypothesis, proposed by Lovelock and Margulis (1974), posits that living organisms interact with their inorganic surroundings to form a synergistic and self-regulating system that maintains the conditions for life on Earth. This theory underscores the intrinsic value of natural systems, emphasising that the health of the entire biosphere is crucial for the survival of its parts. Similarly, deep ecology, advocated by Naess (1973) and further developed by thinkers such as Devall and Sessions (2001), promotes an eccentric view that recognises the inherent worth of all living beings and the interconnectedness of all life forms and the environment. These perspectives highlight the importance of viewing environmental value beyond mere utility, acknowledging the intrinsic and relational values that are essential for the well-being of ecosystems and human societies.

From an assessment perspective, the multi-capitals assessment approach advocated by the Capitals Coalition (2024) offers a comprehensive framework for evaluating the interrelated aspects of the environment, humans, society, and the economy. This approach considers natural, social, human, and produced capitals, ensuring that the impacts and dependencies on various forms of capital are measured and valued. By integrating these different capitals, decision makers can develop more holistic and sustainable strategies that recognise the intrinsic value of natural systems and their critical role in supporting human and economic activities. This method aligns well with the principles of the Gaia hypothesis and deep ecology, as it promotes a balanced and interconnected view of environmental value.

The concepts of doughnut economics, introduced by Raworth (2012), provide another valuable framework for assessing environmental value at a system level. Doughnut economics visualises a safe and just space for humanity, balancing the needs of all within the planet's ecological limits. This model emphasises the imperative of creating economies that are regenerative and distributive by design, ensuring that human activities do not overshoot environmental boundaries while meeting social foundations. By focusing on both ecological ceilings and social foundations, doughnut economics offers a practical approach to achieving sustainability that aligns with the intrinsic and relational values highlighted by the Gaia hypothesis and deep ecology.

While both strong and weak sustainability theories engage with the instrumental value of natural systems, they differ in their approach to resource management and the substitutability of natural capital. Weak sustainability relies on technological solutions and human ingenuity to mitigate environmental impacts, whereas strong sustainability emphasises the preservation and sustainable management of natural resources. In contrast, the Gaia hypothesis and deep ecology focus on the intrinsic value of natural systems, advocating for a holistic and interconnected view of the environment. These theories emphasise the importance of maintaining the health and stability of ecosystems, recognising that human health and well-being are deeply intertwined with the health and well-being of the natural world. They call for a fundamental shift in human attitudes and behaviours towards nature, promoting a more respectful and harmonious relationship with the natural environment.

## 2.2 The environmental impact of civil infrastructure

Environmental value has a direct relationship with civil infrastructure, which supports the natural environment's ability to sustain human life through its resources and provides conditions for economic prosperity. In addition, processes that reduce environmental damage (e.g. renewable energy sources, water collection and distribution, and food production and distribution) require civil infrastructure (Frischmann, 2012). In other words, civil infrastructure has a reciprocal relationship with environmental value, as it relies on the environment for materials, energy, and water while also being necessary for protecting the environment.

This research defines environmental value in this context as the benefits derived from infrastructure systems that contribute to environmental sustainability and overall health and well-being. This concept encompasses several factors, including the reduction of carbon dioxide ($CO_2$) emissions, minimisation of pollution, and the promotion of resource efficiency. Environmental value is not an isolated metric but is deeply interconnected with economic and social values. For example, sustainable transport infrastructure can lead to significant environmental benefits by reducing greenhouse gas (GHG) emissions and improving air quality, which in turn enhances public health and reduces healthcare costs.

Following this approach, we suggest that instrumental and relational value form part of social value and should be studied separately (out of the scope of this study). We highlight the importance of a holistic approach to infrastructure development, where environmental considerations are integrated into the decision making process alongside economic and social factors. By recognising and quantifying these interdependencies, policymakers can develop more robust and sustainable business models for infrastructure projects. The effectiveness of a policy should be evaluated not only in terms of emissions but also its impact on society and the urban environment (Triantafyllopoulos, 2024). This approach ensures that investments not only provide economic returns but also contribute to the long-term sustainability and resilience of communities. By balancing these values, infrastructure systems can better serve the needs of society while preserving the environment for future generations.

A review of relevant literature revealed over 600 publications on the environmental impact of the construction industry and its projects. These studies range from simple to complex analyses of parameters for estimating the environmental impact and show that materials can account for up to 80%–90% of the total structure's embodied GHG emissions (D'Amico and Pomponi, 2018; Kang *et al*., 2015; Zhang and Wang, 2016).

Some studies present simple comparative analyses of environmental impact factors for typical buildings or structures (see Asif *et al*., 2017; Atmaca, 2017; Blok *et al*., 2020; Ding and Forsythe, 2013; Park *et al*., 2014; Puskas and Moga, 2015), while others focus on specific types of structures (see Connolly *et al*., 2018; Lolli *et al*., 2019; Niu and Fink, 2019; Robertson *et al*., 2012) or road and pavement infrastructure (see Balieu *et al*., 2019; Park and Kim, 2019; Said and Al-Qadi, 2019). These analyses are based on a structural analysis of the system of interest without developing new models or theories. For example, Park *et al*. (2014) compared different construction methods and smart frame applications in tall buildings. Robati *et al*. (2019) and Wang *et al*. (2020) used statistical methods such as boxplots to analyse the uncertainty in material quantity calculations. Hodková *et al*. (2012) highlighted that inappropriate use of LCA data could lead to unreliable results, suggesting that initial environmental impact assessments should avoid using LCA databases to reduce uncertainty. The environmental impact can be transformed into emissions using any database after the structural analysis is complete. Statistical analysis of numerous civil projects should provide a range of values to cover the uncertainty of the results (Robati *et al*., 2019; Wang *et al*., 2020). Thus, the optimal way to study the environmental impact of civil infrastructure is to maintain a database of factors impacting the natural environment for numerous construction projects, transforming these factors into environmental value as late as possible in the study process.

Other studies compared the environmental impact of different construction options for civil infrastructure. For instance, Balieu *et al*. (2019) compared different types of electrified road infrastructures, Langston *et al*. (2018) compared refurbished with new projects, and Lemma *et al*. (2020) and Meil *et al*. (2006) compared different designs for the same project. Grant and Ries (2013) studied the life expectancy and replacement of different elements in existing structures. These comparisons did not involve modelling but discussed how different choices affect environmental impact without numerical representation. While comparing construction options is straightforward, numerical modelling, which relies on quantification of the dependency between environmental impact and the influencing parameters, is challenging since often the variables cannot be quantified accurately and/or with certainty.

Establishing the necessary parameters for civil infrastructure usually requires a high level of detail. De Wolf *et al*. (2016, 2020) combined different parameters with embodied carbon coefficients to develop a database estimating global warming potential. The parameters studied were both qualitative (e.g. type of structure, main structural material, rating scheme certification, region, or country) and quantitative (e.g. size by floor area, height by number of floors, number of occupants, and span). Some authors decomposed infrastructure into structural components and estimated emissions using geometric data combined with loading analyses (Collings, 2006; Du and Karoumi, 2013, 2014). The geometric characteristics and inputs affect the material quantities and emissions generated. Other authors studied different structural components, such as beams, columns, walls, and slabs, suggesting optimisation systems to reduce construction materials and $CO_2$ emissions (Hong *et al*., 2010; Lagaros, 2018). Shafiq *et al*. (2015) examined how $CO_2$ emissions from concrete and steel in buildings are affected by the dimensions of structural elements.

A holistic model that includes the entire system of interest is a good approach. Such models can be for simple structures (e.g. water tanks, Sanjuan-Delmás *et al*., 2015) or more complex structures (D'Amico and Pomponi, 2018; Shafiq *et al*., 2015). Sanjuan-Delmás *et al*. (2015) studied how concrete and steel quantities are affected by volume, diameter, height, wall thickness, shape, and material reinforcement, presenting results on a Cartesian coordinate system. D'Amico and Pomponi (2018) created a complex sustainability tool to optimise steel quantities in complex constructions, presenting results by changing one specific parameter while holding other parameters constant, yet this omitted consideration of other potentially important factors (such as temperature, surface area, thickness, and thermal properties): being comprehensive is problematic.

Another approach is the integration of LCA and the Material Circularity Index (MCI), which provides a comprehensive framework for evaluating the combined influences of environmental performance and circularity (Rigamonti and Mancini, 2021). LCA quantifies environmental impacts throughout a product's life cycle, while MCI measures adherence to circular economy principles such as reuse and recycling (Dervishaj and Gudmundsson, 2024). This combined approach helps identify trade-offs between impacts and circularity, supporting balanced strategies and the development of regulations that promote sustainability. However, this approach is challenging to be used on infrastructure systems due to their size.

Based on the above discussion, loadings, geometric parameters, and material quantities of any structure within the infrastructure system must be established to estimate emissions correctly. Combining the reasoning of De Wolf *et al.* (2016, 2020), D'Amico and Pomponi (2018), and Tecchio *et al.* (2019a, 2019b), it is clear that studying the environmental impact of a civil infrastructure system with many projects requires a vast number of parameters for each structure, making the bottom-up approach impractical.

## 3. Methodology

### 3.1 Methods and assumptions

The four major stages within LCA, as certified and supported by the requirements established in the standards of the International Organization for Standardization (ISO), are '(1) goal and scope definition, (2) life cycle inventory (LCI), (3) life cycle impact assessment (LCIA), and (4) life cycle interpretation' (ISO, 2006: p. 2). The imprimatur of ISO contributes to their acceptance by the international community and other stakeholders (Rebitzer *et al.*, 2004).

According to Olugbenga *et al.* (2019), there are five types of LCA methods (see the Video with the five methods online and Figure 1).

- Process-based LCA: this is a bottom-up methodology performed by mapping and characterising 'all processes associated with all life cycle phases of a project' (Jones *et al.*, 2017).
- Hybrid LCA: this method incorporates both top-down economic input–output analysis-based (sector-by-sector wider analysis) and process-based LCA (Chester and Horvath, 2010) to address data gaps when data are available only for part of the process or to expand the boundaries of analysis (García de Soto *et al.*, 2017; Jones *et al.*, 2017).
- Pseudo LCA: this method is based on a mix of primary data and data from literature to calculate GHG emissions. Where system data are not readily available, simplified and parametric LCA approaches are adopted (Bueno *et al.*, 2017; Westin and Kågeson, 2012).
- Simplified LCA: this approach compares the environmental impact of infrastructure with the condition of no infrastructure within a given area (Bueno *et al.*, 2017).
- Parametric LCA: in this method, specific system parameters are statistically modelled to calculate emissions associated with the system.

The two methods with the least uncertainty in assessing environmental impacts will be discussed in the following section: process-based life cycle assessments and a hybrid LCA tool, EE MRIO tables (Figure 1).

### 3.2 Process-based life cycle assessment

A process-based LCA is a method for quantifying the environmental impacts of all processes associated with an existing or potential product or service throughout each stage of its life (Finkbeiner *et al.*, 1998; Jones *et al.*, 2017). This method is the initial approach to conducting a life cycle evaluation of a good, service, or system, focusing on a scientific analysis of the inputs (material and energy balance) and outputs (emissions and wastes; Finkbeiner *et al.*, 2006). The process-based LCA process is a detailed methodology for assessing the environmental impacts of a specific infrastructure system (Ayres, 1995; Trunzo *et al.*, 2019). It is considered a 'bottom-up' method, which is inherently attractive, yet because it quantifies the environmental impacts for each process in all life cycle phases of a project (Olugbenga *et al.*, 2019), by its nature, it is challenging to evaluate infrastructure as a whole. To assess the performance of the civil infrastructure of a society, LCA should be applied to individual infrastructure projects, products, and services and then scaled up to the level (i.e. scale) of interest. Therefore, detailed information is required on all materials and energy used, their supply chains, and the operation, maintenance, and use of every piece of infrastructure being analysed. More specifically, performing a process-based LCA at the level of interest in this research requires extensive information on the extraction, transportation, and assembly of every material used on each site, as well as the recycling or disposal of each component of an infrastructure project at the end of its life.

This method enables a high level of environmental detail to be obtained if the analysis focuses on single products or projects (Beylot *et al.*, 2020). However, many obstacles arise during its application that can restrict the evaluation of impacts. First, due to circular economy practices now more routinely adopted in the pursuit of enhancing environmental value, this method requires many assumptions and decisions to define the goal and scope of the analysis, which can create limitations on the number of components (materials, processes, or flows) and sub-emissions that can be practically considered (Ayres, 1995; Beylot *et al.*, 2020). This makes a comprehensive process-based LCA a very complicated and time-consuming process (Beylot *et al.*, 2020).

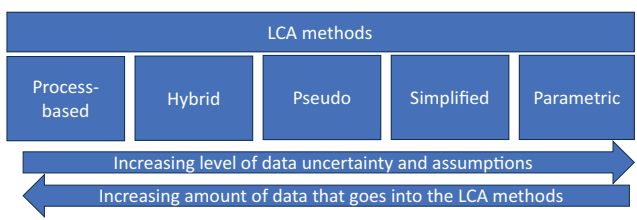

**Figure 1.** LCA methods ranking based on data requirements and uncertainty (Olugbenga *et al.*, 2019)

Second, this method requires extensive primary data from each component within the civil infrastructure. Therefore, if there is insufficient or unreliable data, the analysis will involve an increased level of complexity and uncertainty in determining all the input and output elements, thus reducing the method's utility (Ayres, 1995; Beylot *et al.*, 2020; Trunzo *et al.*, 2019). The data required for this method comes from both public databases and customers. Public databases provide inventory information of specific materials, energy, and processes (Finkbeiner *et al.*, 2006; Lee and Inaba, 2004), while customer data are collected from surveys and manufacturer's assumptions that provide insight into a specific product (Finkbeiner *et al.*, 2006; Lee and Inaba, 2004). Because this data comes from customer's assumptions, it is often 'proprietary, unpublished or confidential', and therefore cannot be verified with other credible databases (Ayres, 1995: p. 8). The larger the system analysis, the greater the data deriving from multiple sources, making the result prone to overstatement (Ayres, 1995). Third, the process-based LCA ignores real human behaviour and, consequently, the economic market and activities related to global commerce (Gutowski, 2018).

Based on the above, the process-based LCA method is appropriate for detailed analysis of specific, single and small-scale infrastructure projects where socio-economic factors are not required to be evaluated. This challenge was one of the main reasons the process-based method was not used in this research, as the scope of the research is to estimate the environmental impact of an entire region (UK) and infrastructure type/sector (transport, energy, water, waste, and communication).

### 3.3 EXIOBASE input–output tables

EE MRIO tables are tools for analysing the environmental impact of economic activities, including infrastructure provision, and their interdependencies across multiple sectors (Stadler *et al.*, 2018). EXIOBASE 3 is an EE MRIO database that combines estimates of the quantity of products supplied and used within different sectors of the economy (input–output tables) with estimates for aggregated emissions to the environment. An input–output table is a tool used in economics to represent the flow of goods and services between different sectors of an economy (Avelino *et al.*, 2021; Langarita and Cazcarro, 2022). EXIOBASE is a 'global multi-regional input–output database' with a high level of sector disaggregation and economic foundation (Stadler *et al.*, 2018). This database has been developed in three different versions through specific projects: the EXIOPOL project developed EXIOBASE1 (2000), the CREEA project created EXIOBASE2 (2007), and the DESIRE project built the latest version, EXIOBASE3 (1995–2011; Tukker *et al.*, 2018). EXIOBASE combines two forms of analysis: monetary (euros) and supply use (tonnes and terajoules, among others) using a range of statistical models (Merciai and Schmidt, 2018). The EE MRIO tables method combines elements of hybrid LCA with parametric LCA methods, as it is based on economic factors and a range of statistical models. An EE MRIO table requires the corresponding input–output table for estimating environmental impacts related to the consumption of products (Tukker *et al.*, 2018; Van Roekel *et al.*, 2017).

EXIOBASE 3 describes the environmental impacts of complex global and cross-sectoral relationships and can therefore inform policymakers regarding the use of resources and the discharge of emissions to the environment (Stadler *et al.*, 2018). EXIOBASE 3 captures both material and service exchanges between sectors (e.g. infrastructure). This tool is important for national and regional decision makers due to its high level of aggregation and coverage of environmental impacts. Therefore, EXIOBASE 3 allows decision makers to take system-wide decisions on a large scale.

EXIOBASE demonstrates a high level of consistency in the macroeconomic sector and international data sources by allowing the estimation of emission factors associated with specific consumption activities within and outside of any particular country (Stadler *et al.*, 2018; Tukker *et al.*, 2018).

However, there are some limitations. First, the latest version of this technique aims to cover global data, but only detailed information from 43 out of 195 countries is included, grouping the remaining countries into five world regions, which may lead to overestimation (Yang *et al.*, 2017). Second, its time series is built for 1995–2011, thus opening up several analytical options, including time series analysis and structural decompositions restricted to these years (Tukker *et al.*, 2018). Third, the classification of emissions requires assumptions in which several elementary flows are not reported in the environmental extensions (Beylot *et al.*, 2020). This prevents the quantification of the entirety of ecological impacts for several impact categories and indicates that this method achieves a low level of detail or approximate sector resolution (Beylot *et al.*, 2020; Suh and Nakamura, 2007). This condition may influence the assessment results and necessitates a sensitivity analysis (Beylot *et al.*, 2020).

Therefore, EXIOBASE 3 is a very powerful tool for decision makers looking to maximise environmental value. However, EXIOBASE 3 cannot provide decision makers with a level of detail sufficient to determine whether one single infrastructure project is better than another in terms of environmental value. Estimating the environmental value of a single infrastructure project using the EXIOBASE 3 would require a very complex top-down analysis.

### 3.4 Critical analysis of methodologies

LCA and EXIOBASE tables capture the environmental value of civil infrastructure in different ways. According to Chan *et al.* (2016), environmental value is defined as the intrinsic, instrumental, and relational value found within nature. In other words, environmental value is a multi-faceted concept, and there is merit in analysing the strengths and weaknesses of both LCA and EXIOBASE tables for each facet—instrumental, intrinsic and relational—of environmental value individually to gain a comprehensive understanding of the overall results, as they relate to civil infrastructure.

Neither method can be said to be superior in capturing nature's intrinsic value, as this depends on the goals and the scale of environmental interventions (Chan *et al.*, 2016). Most of the environmental impacts (and hence value gained or lost) assessed by EXIOBASE 3 and LCA pertain to the instrumental value of the environment. LCAs are useful in identifying whether investment in a specific project or product will affect any of the current 'uses of ecosystem goods' (Daily, 1997). However, when assessed in isolation, an infrastructure project may not cause a substantial environmental impact, whereas several similar projects in aggregate could result in larger consequences. For this reason, EE MRIO tables are more significant in protecting or enhancing the environment's instrumental value. Governments need to manage shared resources among their citizens and industries to ensure that people are served fairly by civil infrastructure, industry is supported economically, and the environmental value of the area is protected (Gómez-Baggethun and Ruiz-Pérez, 2011). These types of decisions can be made with support from EXIOBASE 3.

Neither process-based LCA nor EXIOBASE 3 captures the relational value of the environment. People, cultures, and societies have different relationships with many aspects of nature (Chan *et al.*, 2016). These relationships are challenging to quantify and cannot be measured by either process-based LCAs or EXIOBASE 3. Civil infrastructure can either contribute to or detract from these relationships. For example, when placed in an inappropriate location, infrastructure can permanently scar culturally significant land or damage political relationships (Dona and Singh, 2017).

In addition to Chan's approach, both process-based LCA and EXIOBASE are methods used to quantify environmental footprints using standard mathematical functions (Crawford *et al.*, 2018). Nevertheless, the emphasis of each method will vary depending on the goal, assumptions, interest, and components such as resolution granularity and unit measurement (Beylot *et al.*, 2020; Castellani *et al.*, 2019). Table 1 shows the most notable differences between both approaches.

The above discussion leads to a working hypothesis that process-based LCA is appropriate for small-scale physical civil infrastructures without functionality analysis and within limited and restricted boundaries. Furthermore, this process is suitable if time and effort are invested in finding accurate data for each element within the system, including information provided by industry manufacturers, to ensure the accuracy of the environmental assessment of infrastructure. This method is oriented towards engineers and mid-level decision makers. In contrast, the authors consider EXIOBASE suitable for macro-scale civil infrastructure assessment in an overall context, including physical characteristics, components, and social and economic activities within and outside of the infrastructure's geographical location. EXIOBASE is oriented towards government authorities and international agencies that make high-level decisions. However, both methods should be combined when small-scale civil infrastructure or specific material must be evaluated to estimate environmental impacts that have socio-economic interdependencies. The combination of both will capture missing areas of consumption and provide better and more understandable results.

### 3.5    Summing up the research methodology

Environmental value, and its protection and enhancement, is a complex issue that affects all countries, industries, and economic activities. It encompasses the intrinsic, instrumental, and relational value of nature. Both traditional process-based LCA and EE MRIO tables are valuable tools for evaluating environmental impacts at different scales, with process-based LCA being particularly useful for making specific decisions about products or projects. However, EXIOBASE 3 is more suitable for addressing environmental issues (identifying and characterising environmental impacts and assessing environmental value gained or lost) associated with large-scale interventions and is therefore a better resource for policymakers, who seek to implement significant changes and preserve environmental value both domestically and in their trading partner countries.

Table 1 Comparison of process-based LCA and EXIOBASE

| **Process-based LCA** | **EXIOBASE** |
|---|---|
| Simple methodology (Ayres, 1995) | High level of sector disaggregation (Stadler *et al.*, 2018) |
| Time and effort-consume, but a high level of environmental detail of single products (Beylot *et al.*, 2020) | Environment accounts, micro and macro-scale scenarios (Merciai & Schmidt, 2018) |
| Limited to the areas of consumption and materials (Castellani *et al.*, 2019) | Many products and consumption areas are included in the same database (EXIOBASE Consortium, 2015) |
| ▪ Open database source (Finkbeiner *et al.*, 2006) | ▪ Input–output table (MR-IOT); monetary (Beylot *et al.*, 2020; Tukker *et al.*, 2018) |
| ▪ Private sources from industries (sometimes confidential) (Finkbeiner *et al.*, 2006) | ▪ Multi-regional environmentally extended supply-use table (MR-SUT); supply-use (Beylot *et al.*, 2020; Tukker *et al.*, 2018) |
| ▪ No socio-economic data (Gutowski, 2018) | ▪ Process-based LCA data; coefficients and statistics (Beylot *et al.*, 2020; Tukker *et al.*, 2018) |

## 4. Interpretation, analysis, and discussion

The main challenge in applying these methods is how to track dependencies. Dependencies were studied inductively by examining the correlation between each type of value of the different types of infrastructures, noting that correlation between two variables does not necessarily imply causality (Field, 2009: pp. 619–620). Thus, while two variables can certainly be related with causality, making this assumption and using a superficial interpretation together with the correlation may lead to incorrect conclusions. Accordingly, a causal relationship (interdependence) between two correlated variables was verified with a rational assumption, using the Pearson correlation coefficient for (Field, 2009) this study:

$$r = \frac{s_{xy}}{s_x \cdot s_y} = \frac{\sum_{i=1}^{v}(x_i - \bar{x}) \cdot (y_i - \bar{y})}{\sqrt{\sum_{i=1}^{v}(x_i - \bar{x})^2 \sum_{i=1}^{v}(y_i - \bar{y})^2}}$$

where:

- If $-0.3 < r < 0.3$, there is no linear correlation.
- If $-0.5 < r \leq -0.3$ or $0.3 \leq r < 0.5$, there is a weak linear correlation.
- If $-0.7 < r \leq -0.5$ or $0.5 \leq r < 0.7$, there is a medium linear correlation.
- If $-0.8 < r \leq -0.7$ or $0.7 \leq r < 0.8$, there is a strong linear correlation.
- If $-1 < r \leq -0.8$ or $0.8 \leq r < 1$, there is a very strong linear correlation.
- If $r = \pm 1$, there is a perfect linear correlation.

For this, the existence of linear correlation, meaning $r$ equals more than 0.3 or less than $-0.3$, and the size of the correlation (weak, medium, strong, or very strong) were important.

The interplay between transport and energy emphasises the environmental benefits of adopting advancements such as electric vehicles powered by renewable energy and implementing integrated policies targeting both sectors (Cartone *et al*., 2021; Wei *et al*., 2021). Improved waste management, such as recycling automotive materials, can reduce energy use and emissions (Lee *et al*., 2024). The transport and water sectors are closely linked, with shipping activities affecting water quality and water availability influencing transport operations; innovative technologies and sustainable planning can mitigate these impacts (David and Gollasch, 2015; Makkonen and Inkinen, 2021). In addition, integrating ICT, such as intelligent transportation systems and IoT technologies, enhances transport efficiency, reduces emissions, and supports resilience to climate change (Di Martino *et al*., 2017).

The EXIOBASE 3 database includes 85 types of emissions for both water and air pollution (Stadler *et al*., 2018). EXIOBASE 3 includes EE MRIO tables linked with the economic input–output tables of each country (Stadler *et al*., 2018). The range of emissions considered was, inductively, reduced to those relevant to the focus of this study to make the development of the theory (deduction) manageable—the emissions herein are studied for water and waste pollution, while the sectors of interest are transport, energy, water, waste, and communication. It was checked by which input–output group (IOG) each emission is produced to determine its relevance to this study and, because of its (environmental) focus, the separation of the sectors/IOGs for analysis differs from the separation of sectors/IOGs done in the economic input–output tables.

The EXIOBASE 3 database provides the empirical data used to conclude which types of emissions are produced by the transport, energy, water, waste, and communication sectors. The theory is then derived from the resultant observations, as induction commands (Ghauri and Grønhaug, 2010; May, 2011).

The transport sector includes the IOGs of transportation (economic approach): (*a*) Transport via Railways; (*b*) Other land transport; (*c*) Transport via pipelines; (*d*) Sea and coastal water transport; (*e*) Inland water transport; (*f*) Air transport and the IGOs of manufacturing of vehicles and transport related services (engineering approach); (*g*) Manufacture of motor vehicles, trailers and semi-trailers; (*h*) Manufacture of other transport equipment; (*i*) Sale, maintenance, repair of motor vehicles, motor vehicles parts, motorcycles, motor cycles parts and accessories; and (*j*) Retail sale of automotive fuel.

The energy sector includes (*a*) Production of electricity by coal, (*b*) Production of electricity by gas, (*c*) Production of electricity by nuclear, (*d*) Production of electricity by hydro, (*e*) Production of electricity by wind, (*f*) Production of electricity by petroleum and other oil derivatives, (*g*) Production of electricity by biomass and waste, (*h*) Production of electricity by solar photovoltaic, (*i*) production of electricity, (*j*) Transmission of electricity, (*k*) Distribution and trade of electricity, and (*l*) Manufacture of gas; distribution of gaseous fuels through mains.

The water sector includes (*a*) Steam and hot water supply and (*b*) Collection, purification and distribution of water.

The communication sector includes (*a*) post and telecommunications.

The waste sector includes (*a*) Incineration of waste: Food, (*b*) Incineration of waste: Paper, (*c*) Incineration of waste: Plastic, (*d*) Incineration of waste: Metals and Inert Materials, (*e*) Incineration of waste: Textiles, (*f*) Incineration of waste: Wood, (*g*) Incineration of waste: Oil/Hazardous waste, (*h*) Biogasification of food waste, including land application, (*i*) Biogasification of paper, including land application, (*j*) Biogasification of sewage sludge, including land application, (*k*) Composting of food waste, including land application, (*l*) Composting of paper and wood, including land application, (*m*) Waste water treatment: Food, (*n*) Waste water treatment:

Other, (*o*) Landfill of waste: Food, (*p*) Landfill of waste: Paper, (*q*) Landfill of waste: Plastic, (*r*) Landfill of waste: Inert/metal/hazardous, (*s*) Landfill of waste: Textiles, and (*t*) Landfill of waste: Wood.

To conclude inductively, 26 emissions were found to be produced by the transport sector based on the empirical data (see Table 2). The next step was to seek a connection with the theory, as induction demands. Of the 26 emissions, 23 are calculated by the engineering combustion model calculation of emissions from road transport (COPERT) (Laou, 2013); two (arsenic (As) and mercury (Hg)) are produced only by water transport, and the last (non-methane volatile organic compound(NMVOC)) is a non-combustion pollutant (Stadler *et al*., 2018).

### 4.1 Environmental infrastructure interdependencies

Different methodologies were used for calculating the EXIOBASE emissions of each type of industry, meaning that any correlation will be a result of dependency between the different industries. In addition, each country or area has different sizes of industry developed over its lifecycle.

The key challenge is how to cross-correlate the environmental coefficients to determine their relationships. Pearson correlation is a common method to expose the correlation between series, but since the data were calculated by developing time series of detailed EE MRIO tables, the issues of spurious regression or spurious correlation should be eliminated.

Spurious regression was reduced as follows: Stadler *et al*. (2018) removed all the perfectly correlated indicators (14 indicators) a priori, and the remaining 105 indicators, which also showed very high correlations, were reduced using principal component analysis (PCA) and an optimisation methodology based on the PCA results. This approach eliminated the correlation based on the calculating indicators and the within-series dependence. This can be observed even if coefficients from the same group of sectors are compared with Pearson's correlation, as the correlation is not high in every case and is never perfect, despite using similar indicators.

In addition, the research did not use the chronological development of the data (time series), but the regional development (same year and different country). Each country has a different trajectory over time, has different legislation, and is at a different stage of development. This means it is possible to accept the linearity of the data (linear regression analysis) and assume this linear relationship extends to the total world activity, enabling the application of the Pearson correlation method. This is a safe assumption, as the authors of EXIOBASE 3 also use it, and EXIOBASE 3 is widely used by input–output modellers worldwide. The unknown activity rate is estimated for each year by applying linear regression with a constant offset parameter. Mathematically, it is not the ideal method, but it compensates for a small part of the missing data where real values are absent (Stadler *et al.*, 2018: S3–S7). In addition, the input–output models are linear and do not assume increasing or decreasing returns to scale.

Table 2. Emissions produced by the transport sector

| Emission | Type of emissions (European Environment Agency, 2006) | Assessment method (European Environment Agency, 2006) | Literature/theory |
|---|---|---|---|
| $CO_2$ | Fuel-related pollutants | Fuel consumption | COPERT model (Laou, 2013) |
| $CH_4$ | Non-regulated pollutant | Emissions coefficients | COPERT model (Laou, 2013) |
| $N_2O$ | Non-regulated pollutant | Emissions coefficients | COPERT model (Laou, 2013) |
| $SO_x$ | Non-regulated pollutant | Fuel consumption | COPERT model (Laou, 2013) |
| $NO_x$ | Regulated pollutant | Emissions coefficients | COPERT model (Laou, 2013) |
| $NH_3$ | Non-regulated pollutant | Emissions coefficients | COPERT model (Laou, 2013) |
| CO | Regulated pollutant | Emissions coefficients | COPERT model (Laou, 2013) |
| Benzo(a)pyrene | Fuel-related pollutants | Total percentage of volatile organic compound | COPERT model (Laou, 2013) |
| Benzo(b)fluoranthene | Fuel-related pollutants | Total percentage of volatile organic compound | COPERT model (Laou, 2013) |
| Benzo(k)fluoranthene | Fuel-related pollutants | Total percentage of volatile organic compound | COPERT model (Laou, 2013) |
| Indeno(1,2,3-cd)pyrene | Fuel-related pollutants | Total percentage of volatile organic compound | COPERT model (Laou, 2013) |
| PCDD_F[a] | Fuel-related pollutants | Total percentage of volatile organic compound | COPERT model (Laou, 2013) |
| NMVOC[b] | Non-regulated pollutant | Total percentage of volatile organic compound | COPERT model (Laou, 2013) |
| $PM_{10}$ | Regulated pollutant | Emissions coefficients | COPERT model (Laou, 2013) |
| $PM_{2.5}$ | Regulated pollutant | Emissions coefficients | COPERT model (Laou, 2013) |
| TSP[c] | Regulated pollutant | Emissions coefficients | COPERT model (Laou, 2013) |
| As | Heavy metals | n/a | (Stadler *et al*., 2018) |
| Cd | Heavy metals | Fuel consumption | COPERT model (Laou, 2013) |
| Cr | Heavy metals | Fuel consumption | COPERT model (Laou, 2013) |
| Cu | Heavy metals | Fuel consumption | COPERT model (Laou, 2013) |
| Hg | Heavy metals | n/a | (Stadler *et al*., 2018) |
| Ni | Heavy metals | Fuel consumption | COPERT model (Laou, 2013) |
| Pb | Heavy metals | Fuel consumption | COPERT model (Laou, 2013) |
| Se | Heavy metals | Fuel consumption | COPERT model (Laou, 2013) |
| Zn | Heavy metals | Fuel consumption | COPERT model (Laou, 2013) |
| NMVOC[b] (non-combustion) | Non-regulated pollutant | n/a | (Stadler *et al*., 2018) |

[a] PCDD_F, polychlorinated dibenzodioxins (PCDDs) and polychlorinated dibenzofurans (PCDFs)
[b] NMVOC, non-methane volatile organic compounds
[c] TSP, total suspended particles

To summarise, the IOGs that belong to the same sector group are correlated because a similar methodology was used to estimate the emissions produced. Therefore, these correlations are rejected since it is not possible to establish whether they are correlated because of the estimation methodology, meaning that the results are biased.

Tables with the calculated correlations between the sectors of interest for each pollutant were created. Table 3 is an example table for $CO_2$. The sectors within the same group were highlighted and removed (see Column 3 in Table 3). Then, the number of missing data points was checked, and if the missing data exceeded 10% of the total, the connection was also removed (see Column 2 in Table 3). For example, EnergyNuclear may be dependent on other sectors, but since many of the areas studied do not have this type of energy, it is not possible to include the connection.

The IOGs glossary of the first column of Table 3 is as follows.

- TRail is the 'Transport via Railways' IOG of EXIOBASE
- TOther is the 'Other land transport' IOG of EXIOBASE
- TPipelines is the 'Transport via pipelines' IOG of EXIOBASE
- TSea is the 'Sea and coastal water transport' IOG of EXIOBASE
- TManufMotor is the 'Manufacture of motor vehicles, trailers, and semi-trailers' IOG of EXIOBASE
- TManufOther is the 'Manufacture of other transport equipment' IOG of EXIOBASE
- TAir is the 'Air transport' IOG of EXIOBASE
- TSaleFuel is the 'Sale, maintenance, repair of motor vehicles, motor vehicles parts, motorcycles, motorcycles parts and accessories' IOG of EXIOBASE
- EnergyCoal is the 'Production of electricity by coal' IOG of EXIOBASE
- EnergyGas is the 'Production of electricity by gas' IOG of EXIOBASE
- EnergyNuclear is the 'Production of electricity by nuclear' IOG of EXIOBASE
- EnergyWind is the 'Production of electricity by wind' IOG of EXIOBASE
- EnergyHydro is the 'Production of electricity by hydro' IOG of EXIOBASE
- EnergySolar1 is the 'Production of electricity by solar photovoltaic' IOG of EXIOBASE
- EnergySolar2 is the 'Production of electricity by solar thermal' IOG of EXIOBASE
- EnergyTransm is the 'Transmission of electricity' IOG of EXIOBASE
- Energynec is the 'Production of electricity not elsewhere classified' IOG of EXIOBASE
- EnergyPetrol is the 'Production of electricity by petroleum and other oil derivatives' IOG of EXIOBASE
- EnergyBiomass is the 'Production of electricity by biomass and waste' IOG of EXIOBASE
- EnergyOcean is the 'Production of electricity by tide, wave, ocean' IOG of EXIOBASE
- EnergyGeoth is the 'Production of electricity by Geothermal' IOG of EXIOBASE
- EnergyTransm is the 'Transmission of electricity' IOG of EXIOBASE
- EnergyDistrib is the 'Distribution and trade of electricity' IOG of EXIOBASE
- Communic is the 'Post and telecommunications' IOG of EXIOBASE
- NWaterSupply the 'Steam and hot water supply' IOG of EXIOBASE
- NWaterDistrib is the 'Collection, purification, and distribution of water' IOG of EXIOBASE
- WIFood is the 'Incineration of waste: Food' IOG of EXIOBASE
- WIPaper is the 'Incineration of waste: Paper' IOG of EXIOBASE
- WIMetal is the 'Incineration of waste: Metals and Inert Materials' IOG of EXIOBASE
- WITextile is the 'Incineration of waste: Textiles' IOG of EXIOBASE
- WIWood is the 'Incineration of waste: Wood' IOG of EXIOBASE
- WIOil is the 'Incineration of waste: Oil/Hazardous waste' IOG of EXIOBASE
- WWFood is the 'Wastewater treatment, food' IOG of EXIOBASE
- WWOther is the 'Wastewater treatment, other' IOG of EXIOBASE
- WCFood is the 'Composting of food waste, incl. land application' IOG of EXIOBASE
- WCPaper is the 'Composting of paper and wood, incl. land application' IOG of EXIOBASE
- WBPaper is the 'Biogasification of paper, incl. land application' IOG of EXIOBASE
- WBFood is the 'Biogasification of food waste, incl. land application' IOG of EXIOBASE
- WBSewage is the 'Biogasification of sewage slugde, incl. land application' IOG of EXIOBASE
- WLFood is the 'Landfill of waste: Food' IOG of EXIOBASE
- WLPaper is the 'Landfill of waste: Paper' IOG of EXIOBASE
- WLPlastic is the 'Landfill of waste: Plastic' IOG of EXIOBASE
- WLMetal is the 'Landfill of waste: Inert/Metal/Hazardous' IOG of EXIOBASE
- WLTextile is the 'Landfill of waste: Textiles' IOG of EXIOBASE
- WLWood is the 'Landfill of waste: Wood' IOG of EXIOBASE

Table 3. $CO_2$ emission generation correlation between sectors

| Input–output groups | Correlation—$CO_2$ | Data robust[a] | Type |
| --- | --- | --- | --- |
| WIPlastic–WITextile | 1.000 | x | **W-W** |
| WBFood–WBPaper | 0.999 | X | **W-W** |
| WIMetal–WCFood | 0.999 | x | **W-W** |
| WIMetal–WIOil | 0.993 | x | **W-W** |
| NWaterSupply–WIMetal | 0.993 | x | N-W |
| NWaterSupply–WCFood | 0.992 | x | N-W |
| WLFood–WLTextile | 0.991 | ✓ | **W-W** |
| WLPaper–WLTextile | 0.990 | ✓ | **W-W** |
| WLFood–WLPaper | 0.990 | ✓ | **W-W** |
| WLPlastic–WLWood | 0.957 | ✓ | **W-W** |
| WIOil–WCFood | 0.928 | x | **W-W** |
| NWaterSupply–WLMetal | 0.919 | ✓ | N-W |
| NWaterSupply–WIOil | 0.919 | x | N-W |
| WIMetal–WLMetal | 0.919 | x | **W-W** |
| EnergyTransm–EnergyDistrib | 0.918 | ✓ | **E-E** |
| WCFood–WLMetal | 0.913 | x | **W-W** |
| WLFood–WLWood | 0.893 | ✓ | **W-W** |
| WLPaper–WLWood | 0.888 | x | **W-W** |
| WIOil–WLMetal | 0.887 | x | **W-W** |
| WIFood–WIPaper | 0.874 | x | **W-W** |
| WLTextile–WLWood | 0.872 | ✓ | **W-W** |
| EnergyCoal–EnergyDistrib | 0.846 | ✓ | **E-E** |
| EnergyTransm–NWaterDistrib | 0.841 | ✓ | E-N |
| EnergyCoal–EnergyTransm | 0.828 | ✓ | **E-E** |
| WWFood–WWOther | 0.816 | ✓ | **W-W** |
| TManufMotor–TPipelines | 0.803 | ✓ | T-T |
| EnergyNuclear–EnergyWind | 0.767 | x | **E-E** |
| EnergyDistrib–NWaterDistrib | 0.761 | ✓ | E-N |
| WLPaper–WLPlastic | 0.753 | ✓ | **W-W** |
| WLFood–WLPlastic | 0.750 | ✓ | **W-W** |
| EnergyOcean–WCPaper | 0.731 | x | E-W |
| TSaleFuel–Communic | 0.725 | ✓ | T-C |
| WLPlastic–WLTextile | 0.715 | ✓ | **W-W** |
| EnergyCoal–NWaterDistrib | 0.707 | ✓ | E-N |
| TRail–Communic | 0.617 | ✓ | T-C |
| EnergyPetrol–TAir | 0.598 | ✓ | T-E |
| WWFood–WLFood | 0.591 | ✓ | **W-W** |
| TSea–WIPlastic | 0.591 | x | T-W |
| TSea–WITextile | 0.591 | x | T-W |
| EnergyGeoth–TPipelines | 0.589 | x | E-T |
| WWFood–WLPaper | 0.579 | ✓ | **W-W** |
| WIOil–WWOther | 0.579 | x | **W-W** |
| WWFood–WLTextile | 0.565 | ✓ | **W-W** |
| TManufMotor–EnergyGeoth | 0.556 | ✓ | T-E |
| TOther–TAir | 0.554 | ✓ | **T-T** |
| TManufMotor–TManufOther | 0.553 | ✓ | **T-T** |
| EnergyBiomass–EManufGas | 0.547 | ✓ | E-E |
| EnergySolar2–EnergyGeoth | 0.545 | x | E-E |
| WWOther–WLMetal | 0.538 | ✓ | **W-W** |
| EnergyNuclear–NWaterDistrib | 0.537 | x | E-N |
| EnergyDistrib–TOther | 0.492 | ✓ | E-T |
| TManufOther–TPipelines | 0.489 | ✓ | **T-T** |
| EnergyNuclear–TAir | 0.488 | x | E-T |
| WIFood–WIPlastic | 0.487 | x | **W-W** |
| WIFood–WITextile | 0.487 | x | **W-W** |
| TManufMotor–TRail | 0.484 | ✓ | **T-T** |
| EnergyTransm–TOther | 0.483 | ✓ | E-T |
| EnergyNuclear–EnergyTransm | 0.481 | x | **E-E** |
| EnergyNuclear–TOther | 0.477 | x | E-T |

**Table 3.** Continued

| Input–output groups | Correlation—$CO_2$ | Data robust[a] | Type |
|---|---|---|---|
| NWaterDistrib–TOther | 0.474 | ✓ | N-T |
| EnergyBiomass–TPipelines | 0.468 | ✓ | E-T |
| TManufMotor–EnergyWind | 0.464 | ✓ | T-E |
| EnergyPetrol–TOther | 0.463 | ✓ | T-E |
| Energynec–WWOther | 0.462 | ✓ | E-W |
| EnergyWind–NWaterDistrib | 0.462 | x | E-N |
| EnergyPetrol–EnergyDistrib | 0.475 | ✓ | **E-E** |
| Energynec–WWFood | 0.455 | x | E-W |
| NWaterSupply–WWOther | 0.449 | x | N-W |
| EnergyWind–EnergyTransm | 0.444 | x | **E-E** |
| EnergyCoal–EnegryPetrol | 0.442 | ✓ | **E-E** |
| EnergyWind–TPipelines | 0.440 | x | **E-E** |
| WBFood–WBSewage | 0.434 | x | **W-W** |
| TSea–WIWood | 0.430 | x | T-W |
| WBPaper–WBSewage | 0.430 | x | **W-W** |
| WIMetal–WWOther | 0.428 | x | **W-W** |
| WWFood–WLMetal | 0.421 | x | **W-W** |
| EnergySolar1–EnergySolar2 | 0.420 | x | **E-E** |
| WCFood–WWOther | 0.420 | x | **W-W** |
| EnergyNuclear–EnergyDistrib | 0.410 | x | **E-E** |
| EnergyNuclear–EnegryPetrol | 0.409 | x | **E-E** |
| WIOil–WWFood | 0.407 | x | **W-W** |
| TSaleFuel–WLPaper | 0.407 | x | T-W |
| TRail-TSaleFuel | 0.406 | ✓ | **T-T** |
| EnergyGas–Communic | 0.405 | ✓ | E-C |
| EnergySolar1–TRail | 0.401 | x | E-T |
| TSaleFuel–WLFood | 0.396 | ✓ | E-W |
| TSaleFuel–WLTextile | 0.387 | ✓ | T-W |
| EnergyGas–TRail | 0.385 | ✓ | E-T |
| WBFood–WCPaper | 0.381 | x | **W-W** |
| WWFood–WLWood | 0.381 | x | **W-W** |
| EnergyGas–TWaterLand | 0.378 | ✓ | E-T |
| WBPaper–WCPaper | 0.377 | x | **W-W** |

[a] The connection is included in the analysis if fewer than 10% of the data points used in the correlation are missing (✓); if more than 10% are missing, the connection is removed (x)

The third column of Table 3 shows the environmental dependence between the sectors of transport (T), communication (C), energy (E), water (N), and waste (W), for example T–C means that transport and communication sectors are dependent for the specific IOG. A network diagram was developed to represent the findings (see Figure 2), integrating statistical significance levels as a methodological approach (see Niavis *et al.*, 2024).

The same process was repeated for the other 25 emissions, the results being detailed in Table 4 and represented in Figure 3.

## 5. Concluding discussion

The statistical inference of EXIOBASE demonstrated that there are emission generation interdependencies that can be quantified. It is important to note that the linearity assumptions influence the analysis, yet LCA and EXIOBASE remain widely accepted.

Very many of the emissions generated by energy and land transport sectors are correlated ($CO_2$, methane ($CH_4$), nitrous oxide ($N_2O$), sulfur oxide ($SO_X$), nitrogen oxide ($NO_X$), ammonia ($NH_3$), carbon monoxide (CO), Benzo(a)pyrene, Benzo(b)fluoranthene, Benzo(k)fluoranthene, Indeno(1,2,3-cd)pyrene, NMVOC, particulate matter (PM10): inhalable particles, with diameters that are generally 10 μm and smaller ($PM_{10}$), particulate matter (PM2.5): inhalable particles, with diameters that are generally 2.5 μm and smaller ($PM_{2.5}$), triple superphosphate (TSP), cadmium (Cd), chromium (Cr), copper (Cu), nickel (Ni), lead (Pb), selenium (Se), zinc (Zn), and NMVOC (non-combustion)), while fewer emissions correlate with air transport ($CO_2$, $CH_4$, $SO_X$, $NH_3$, CO, Benzo(k)fluoranthene, Indeno(1,2,3-cd)pyrene, NMVOC, $PM_{10}$, $PM_{2.5}$, TSP, Cd, Cr, Cu, Ni, Pb, Se, Zn), and even fewer still ($CO_2$, $NH_3$, CO, Benzo(b)fluoranthene, Benzo(k)fluoranthene, and TSP) correlate with water transport sectors. In general, many of the emissions generated by land transport correlate with those from the communication sector ($CO_2$, $N_2O$, $SO_X$, $NO_X$, $NH_3$, CO, Benzo(a)pyrene, Benzo(b)fluoranthene, Benzo(k)fluoranthene, NMVOC, $PM_{10}$, $PM_{2.5}$, TSP, Cr, Cu, Pb, Se, and Zn) and water sector ($CO_2$, $NO_X$, $NH_3$, CO, NMVOC, $PM_{10}$, $PM_{2.5}$, TSP, Cd, Cu, Pb, Se, Zn, and NMVOC (non-combustion)). Relatively few of the emissions generated by the waste and land transport sectors are correlated ($CO_2$, $NH_3$, CO, Benzo(a)pyrene, and NMVOC (non-combustion)). Water transport emissions correlate with those from the water ($CO_2$ and $NH_3$), communication ($CO_2$), and waste ($CO_2$, $NH_3$, and Cu) sectors. Finally, there are air transport emissions that correlate with the water ($CO_2$, $NH_3$, Cd, and Pb), communication ($CO_2$, $NO_X$, and CO), and waste ($CO_2$, $SO_X$, and $NH_3$) sectors.

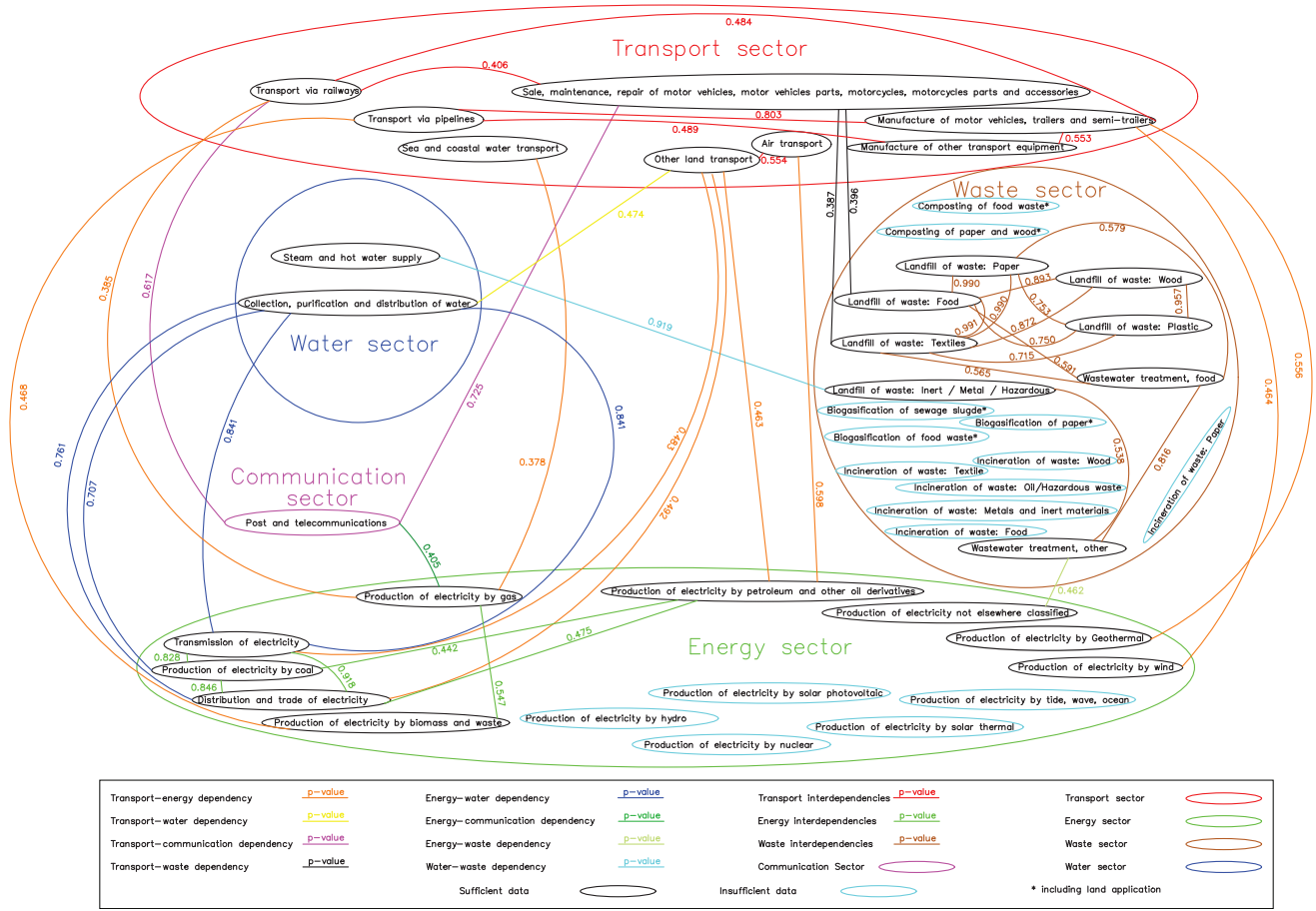

**Figure 2.** Network diagram of $CO_2$ emission generation correlation between sectors

The correlation of emissions between the transport and energy sectors is a vital area of research, as both sectors significantly contribute to these environmental impacts. Although data reveal a weak to moderate dependency, nearly all IOGs within these sectors exhibit at least one correlation with IOGs from the other (the exceptions being two IOGs in the transport sector and one in energy). These findings broadly align with established expectations in the literature but are made specific herein.

The relationship between the transport and waste sectors is complex. Transport directly correlates weakly with the IOGs of food and textile waste landfills while indirectly influencing and being influenced by other waste IOGs that correlate with these two. However, given the absence of direct correlations of transport with other IOGs, drawing definitive conclusions is challenging. Through these indirect interactions, transport also connects with the energy and water sectors by way of the waste sector. Conversely, waste management practices can directly impact the environmental footprint of the transport sector.

The connection between land transport and water resources is relatively weak, limited to a correlation with the steam and hot water supply IOG. This IOG also correlates with the energy sector, creating an indirect interaction between energy and transport.

The interconnection between transport and communication is evident, with medium to strong correlations found in rail transport and the sale, maintenance, and repair of motor vehicles, parts, and accessories IOGs.

Intra-sectoral interdependencies often exhibit strong or very strong correlations. However, water demonstrates a strong to very strong dependency on energy, while communication shows a medium to strong dependency on transport.

In conclusion, the correlations between transport and the other economic infrastructures (energy, waste, water, and communication)

Table 4. Environmental infrastructure interdependencies

| Emission | Transport dependency | Air | Land | Water |
|---|---|---|---|---|
| $CO_2$ | Energy; communication; water; waste | Energy; communication; water; waste | Energy; communication; water; waste | Energy; communication; water; waste |
| $CH_4$ | Energy | Energy | Energy | — |
| $N_2O$ | Energy; communication | — | Energy; communication | — |
| $SO_x$ | Energy; communication; water; waste | Energy; waste | Energy; communication | — |
| $NO_x$ | Energy; communication; water; waste | Communication | Energy; communication; water | — |
| $NH_3$ | Energy; communication; water; waste | Energy; water; waste | Energy; communication; water; waste | Energy; Water; Waste |
| CO | Energy; communication; water; waste | Energy; communication | Energy; communication; water; waste | — |
| Benzo(a)pyrene | Energy; communication; waste | — | Energy; Communication; Waste | — |
| Benzo(b)fluoranthene | Energy; communication; waste | — | Energy; communication | Energy |
| Benzo(k)fluoranthene | Energy; communication | Energy | Energy; communication | Energy |
| Indeno(1,2,3-cd) pyrene | Energy; communication; water | Energy | Energy | — |
| PCDD_F | (Missing values) | (Missing values) | (Missing values) | (Missing values) |
| NMVOC | Energy; communication; water | Energy | Energy; communication; water | — |
| $PM_{10}$ | Energy; communication; water | Energy | Energy; communication; water | — |
| $PM_{2.5}$ | Energy; communication; water | Energy | Energy; communication; water | — |
| TSP | Energy; communication; water | Energy | Energy; communication; water | Energy |
| As | (Missing values) | (Missing values) | (Missing values) | (Missing values) |
| Cd | Energy; water | Energy; water | Energy; water | — |
| Cr | Energy; communication | Energy | Energy; communication | — |
| Cu | Energy; communication; water | Energy | Energy; communication; water | Waste |
| Hg | (Missing values) | (Missing values) | (Missing values) | (Missing values) |
| Ni | Energy | Energy | Energy | — |
| Pb | Energy; communication; water | Energy; water | Energy; communication; water | — |
| Se | Energy; communication; water | Energy | Energy; communication; cater | — |
| Zn | Energy; communication; water | Energy | Energy; communication; water | — |
| NMVOC (non-combustion) | Energy; water; waste | — | Energy; water; waste | — |

underscore the interconnected nature of environmental impacts. These relationships point to opportunities for targeted innovations to significantly reduce emissions. Integrated policies and advanced technologies can amplify these benefits, with methodologies such as Pearson correlation and PCA ensuring reliable and actionable insights. Future research should further explore these interdependencies, utilising advancements in data analytics and modelling to develop holistic strategies for sustainable development.

A more in-depth environmental analysis is required to understand how the production of different pollutants interacts, as these relationships are essentially the result of economic analysis (similar to hybrid LCA methods) based on quantitative data rather than qualitative. Taking this argument further, future research should focus on developing integrated models that combine quantitative and qualitative data to better understand the cause–effect relationships between different pollutants produced by transport infrastructure. This approach can help policymakers make informed decisions on promoting sustainable transport infrastructure. In addition, expanding the scope of studies to include diverse geographical regions will enhance the generalisability of findings and support the development of globally applicable environmental policies.

## Acknowledgements

The authors gratefully acknowledge the University of Birmingham, the University of Illinois at Urbana–Champaign, the University of

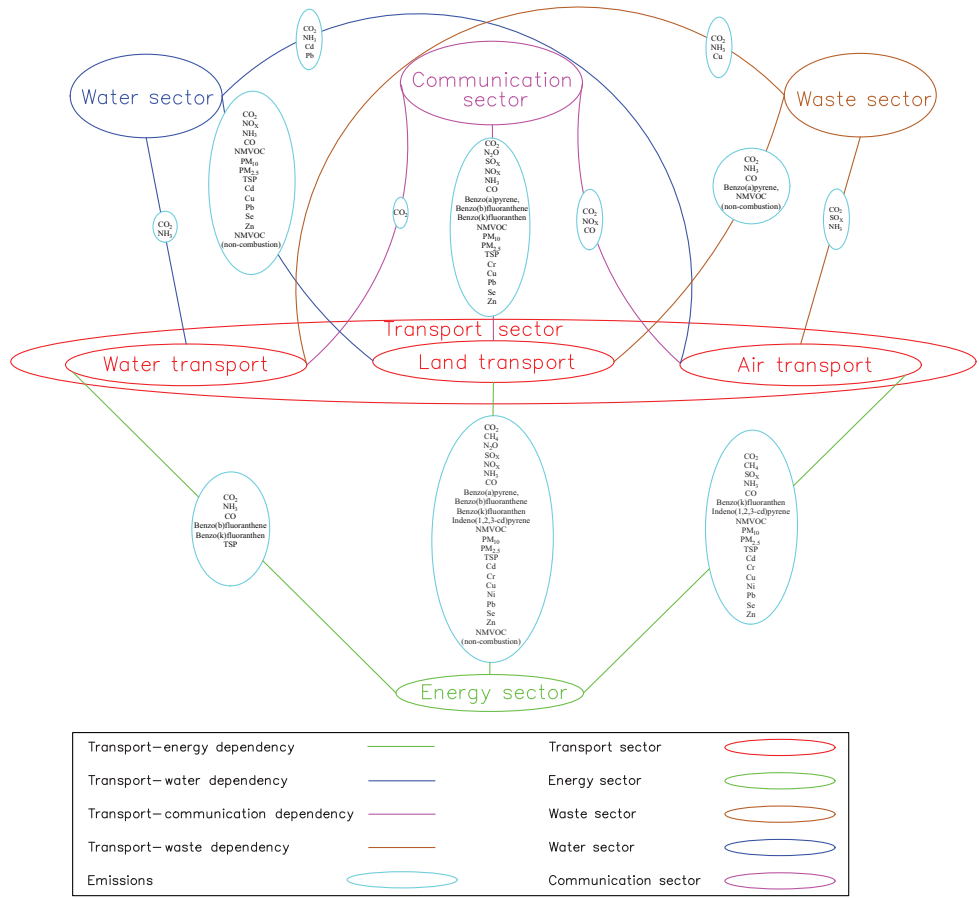

**Figure 3.** Emission dependencies of transport sector

Thessaly and the financial support of the UK Engineering and Physical Sciences Research Council under grant numbers EP/K012398 (iBUILD: Infrastructure Business Models, Valuation and Innovation for Local Delivery), EP/J017698 (Transforming the Engineering of Cities to Deliver Societal and Planetary Wellbeing, known as Liveable Cities), and EP/R017727 (UK Collaboratorium for Research on Infrastructure and Cities). This research informed a newly awarded grant from the Medical Research Council for the Healthy Low-Carbon Transport Hub (Grant No. MR/Z506382).